\documentclass[proceedings, onecolumn]{rmaa}
\usepackage[utf8]{inputenc}
\usepackage{graphicx}
\usepackage{color}
\usepackage{hyperref}
\usepackage{natbib}
\usepackage{url}

\usepackage{subfigure}

\title{DECam-GROWTH search for the faint and distant binary neutron star and neutron star-black hole mergers in O3a}

\author{
Shreya~Anand,\altaffiltext{1}{Division of Physics, Mathematics, and Astronomy, California Institute of Technology, Pasadena, CA 91125, USA}\altaffilmark{1} 
Igor~Andreoni,\altaffilmark{1} 
Daniel~A.~Goldstein,\altaffilmark{1}
Mansi~M.~Kasliwal,\altaffilmark{1}
Tom{\'a}s~Ahumada,\altaffiltext{2}{Department of Astronomy, University of Maryland, College Park, MD 20742, USA} \altaffilmark{2} 
Jennifer~Barnes,\altaffiltext{3}{Columbia Astrophysics Laboratory, Columbia University, New York, NY,10032}\altaffilmark{3} 
Joshua~S.~Bloom,\altaffiltext{4}{Lawrence Berkeley National Laboratory, 1 Cyclotron Road, Berkeley, CA, 94720, USA} \altaffiltext{5}{Department of Astronomy, University of California, Berkeley, 94720, USA}\altaffilmark{4,5} 
Mattia~Bulla,\altaffiltext{6}{Nordita, KTH Royal Institute of Technology and Stockholm University, Roslagstullsbacken 23, SE-106 91 Stockholm, Sweden}
\altaffiltext{7}{The Oskar Klein Centre, Department of Physics, Stockholm University, AlbaNova, SE-106 91 Stockholm, Sweden}\altaffilmark{6,7}
S.~Bradley~Cenko,\altaffiltext{8}{Astrophysics Science Division, NASA Goddard Space Flight Center, MC 661, Greenbelt, MD 20771, USA}\altaffiltext{9}{Joint Space-Science Institute, University of Maryland, College Park, MD 20742, USA}\altaffilmark{8,9}
Jeff~Cooke,\altaffiltext{10}{Australian Research Council Centre of Excellence for Gravitational Wave Discovery (OzGrav), Swinburne University of Technology, Hawthorn, VIC, 3122, Australia}
\altaffiltext{11}{Centre for Astrophysics and Supercomputing, Swinburne University of Technology, Hawthorn, VIC, 3122, Australia}\altaffilmark{10,11}
Michael~W.~Coughlin,\altaffiltext{12}{Department of Physics and Astronomy, University of Minnesota, Minneapolis, Minnesota, 55455}\altaffilmark{12}
Peter~E.~Nugent,\altaffilmark{4,5}
and
Leo P. Singer \altaffilmark{8}
}
\date{February 2020}

\abstract{Synoptic searches for the optical counterpart to a binary neutron star (BNS) or neutron star-black hole (NSBH) merger can pose significant challenges towards the discovery of kilonovae and performing multi-messenger science. In this work, we describe the advantage of a global multi-telescope network towards this end, with a particular focus on the key and complementary role the Dark Energy Camera (DECam) plays in multi-facility follow-up. We describe the Global Relay of Observatories Watching Transients Happen (GROWTH) Target-of-Opportunity (ToO) Marshal, a common web application we built to ingest events, plan observations, search for transient candidates, and retrieve performance summary statistics for all of the telescopes in our network. Our infrastructure enabled us to conduct observations of two events during O3a, S190426c and S190510g. Furthermore, our analysis of deep DECam observations of S190814bv conducted by the DESGW team, and access to a variety of global follow-up facilities allowed us to place meaningful constraints on the parameters of the kilonova and the merging binary. We emphasize the importance of a global telescope network in conjunction with a power telescope like DECam in performing searches for the counterparts to gravitational-wave sources.}

\shortauthor{Anand et al.}
\shorttitle{DECam-GROWTH Follow-ups in O3a}
\begin{document}
\maketitle

\section{Introduction}

The direct detection of gravitational wave (GW) signals and the discovery of their electromagnetic (EM) counterparts opened a new era for multi-messenger astrophysics. The combined discovery of a GW event and associated EM transients can lead to a watershed of science, for example providing insights into heavy-element nucleosynthesis, neutron star (NS) equation of state, relativistic jet physics, stellar evolution, tests of general relativity, and cosmology. In 2017, the identification and follow-up of an EM counterpart to the nearby GW170817 event proved the benefit of multi-wavelength follow-up from radio to gamma-rays (e.g. \citealt{LVC2017, MMA2017}). Prompt optical and near-infrared (IR) observations can lead to the sub-arcsecond localization of a transient called ``kilonova" (KN) or ``macronova" associated with the the GW source. Accurate luminosity distances can then be measured with spectroscopic follow-up of the kilonova and its host galaxy.

The third LIGO-Virgo-KAGRA observing run (O3) started in April 2019 and it is expected to conclude in May 2020. Open alerts are made available to the community upon GW discovery, which enables immediate EM follow-up. Several reasons concur to make rapid follow-up necessary, including i) the optical/IR counterpart, if exists, is expected to fade significantly in hours to days; ii) early detection can enable extensive spectroscopic and photometric multi-wavelength follow-up; 3) even in the case of GW170817, the beginning of the light curve evolution was not well understood and only observations within $<10$ hours can help break the degeneracy between models (eg. \citealt{Ar2018} and references therein).   

O3 has yielded more than 53 events from its start to March 2020, with several BNS and NSBH mergers that may be accompanied by EM counterparts. However, no counterpart to a GW signal has been identified during O3, yet. The search for an optical/IR counterpart to mergers with at least 1 NS is particularly challenging during O3, mainly because of the large sky areas to cover (with a median 90\% integrated probability region of $\sim 3400$deg$^2$, for BNS and NSBH), disjoint skymaps where high-probability lobes are located far apart from each other, and large estimated distances (usually $>200$Mpc). Both galaxy-targeted and synoptic follow-up strategies were used by astronomers during O3. Here we present how the GROWTH collaboration plans and conducts multi-facility follow-up of GW triggers. In particular we describe the custom software that we created for multi-telescope observation scheduling and we present the DECam-GROWTH project, which uses the wide-field Dark Energy Camera for deep, synoptic follow-up of NS mergers.

\section{The GROWTH Network} \label{sec:growth}
The Global Relay of Observatories Watching Transients Happen (GROWTH) is a network of partner institutions and telescopes around the world aimed at obtaining continuous follow-up of various kinds of fast transients. In the context of the search for electromagnetic counterparts to gravitational waves, our network consists of a number of ``search" telescopes capable of conducting galaxy-targeted and synoptic searches for gravitational waves, including the Zwicky Transient Facility (ZTF; \citealt{BeKu2019}), the Dark Energy Camera (DECam; \citealt{FlDi2015}), Palomar Gattini-IR \citep{DeHa2020, MoKa2019}, Kitt Peak EMCCD Demonstrator (KPED; \citealt{CoDe2019}), and GROWTH-India telescope (GIT \footnote{https://sites.google.com/view/growthindia/}). The majority of the other facilities in our network obtain photometric and spectroscopic follow-up of transient candidates identified during the searches within the GW localization.

\subsection{GROWTH Target-of-Opportunity Marshal} \label{sec:growth_marshal}
To facilitate the rapid and coordinated follow-up of transient events localized to large sky regions (such as GRBs, GWs, and neutrinos) as soon as the skymaps are distributed via GCN notice and become visible at our observatory sites, we developed the GROWTH Target-of-Opportunity (ToO) Marshal\footnote{https://github.com/growth-astro/growth-too-marshal} \citep{KaCa2019}. The GROWTH ToO marshal is a Flask-based web application, built on a postgresql database, relying on celery for the management of an asynchronous task queue to handle long-term background jobs. \texttt{ligo.skymap} \footnote{https://github.com/lpsinger/ligo.skymap}, an open-source software for processing and manipulating gravitational-wave localizations, is a key component of the ToO Marshal. The GROWTH ToO Marshal automatically ingests all GCN notices and displays them on the front page; users can navigate to a given event and generate different potential observing plans to either tile the localization or perform galaxy-targeted follow-up.  The backend for scheduling is handled by Gravitational-Wave ElectroMagnetic Optimization (\texttt{gwemopt} \footnote{https://github.com/mcoughlin/gwemopt} \citealt{CoTa2018}), another open-source codebase used to apply algorithms for tiling, time allocation, and scheduling to facilitate planning observations within a skymap.

The ToO marshal's key functionalities include:
\begin{enumerate}
    \item Automatically ingesting GCN notices from GRB, GW, and neutrino alerts
    \item Notifying astronomers (via phone call and text message) when a multi-messenger event, worthy of follow-up, has been detected
    \item Displaying classification and source properties of all ingested events
    \item Facilitating observing plan generation through a variety of customizable settings
    \item Visualizing proposed plan via data tables (containing probability coverage and total time) and planned pointings onto the skymap
    \item Allowing users to directly trigger the telescope queue
    \item Generating tables containing all galaxies falling within the 90\% credible volume
    \item Querying our GROWTH Marshal database \citep{KaCa2019} for saved transient candidates falling within the 2D skymap
    \item Retrieving completed observations and calculating statistics for the GCN (such as probability, coverage, and median depth)
\end{enumerate}

\begin{figure*}[h!]
    \centering
    \includegraphics[width=0.4\textwidth]{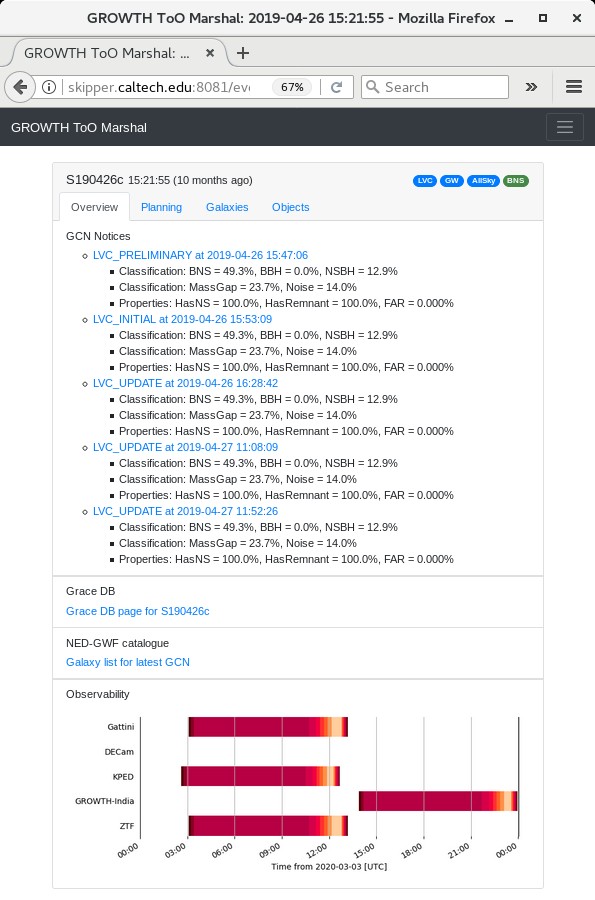}\includegraphics[width=0.4\textwidth]{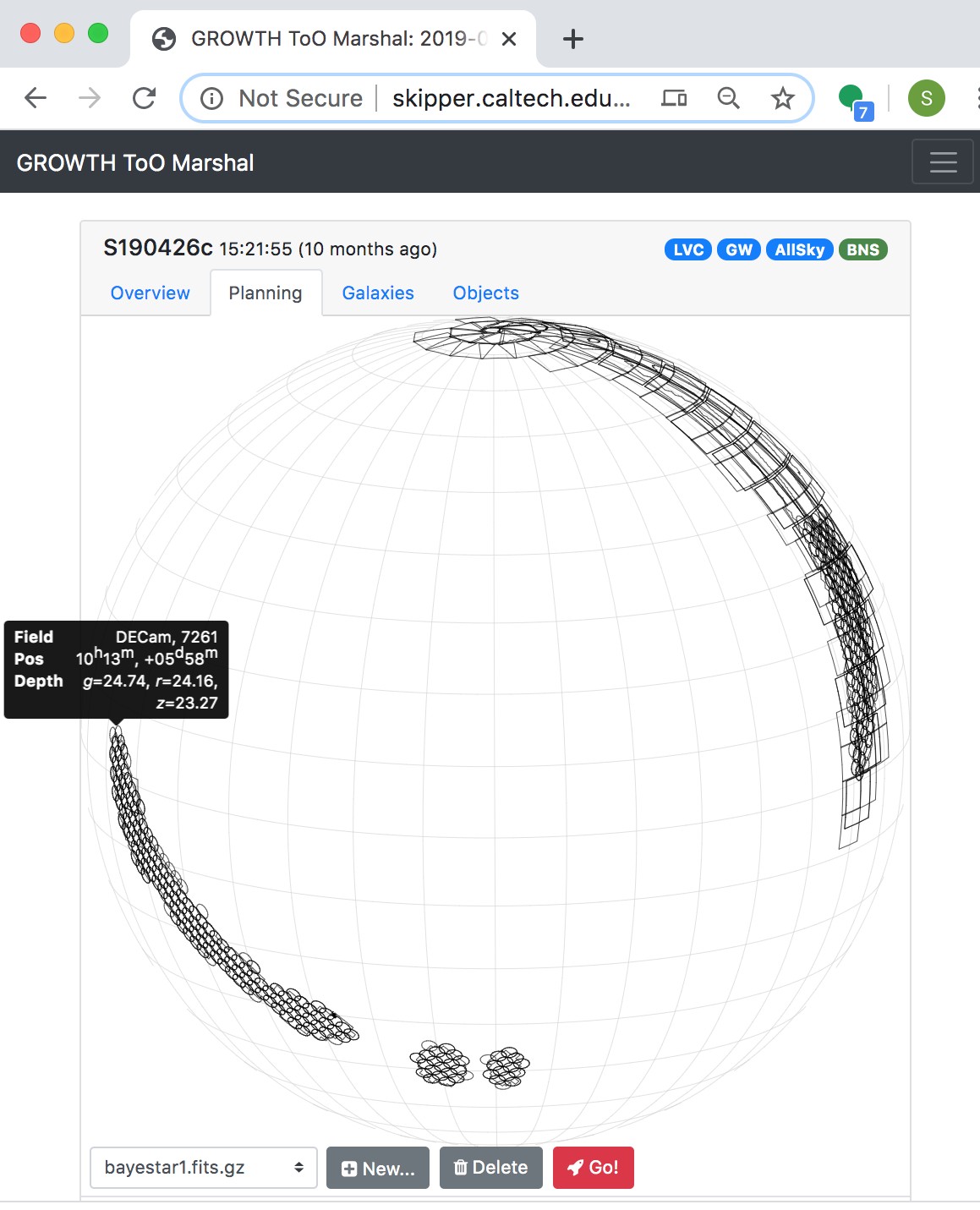}
    \caption{Screenshots from the GROWTH ToO Marshal, displaying the event page, and the plan page. These screenshots feature S190426c, the first event during O3a that we triggered DECam on. Left: the event page, used to display properties of each ingested GCN notices ascribed to S190426c, and to show the observability of the skymap as a function of time from each of our GROWTH search telescopes; Right: disjointed BAYESTAR skymap for S190426c - DECam planned pointings (shown in circles) cover the entire Southern lobe and the lower portion of the Northern lobe, ZTF planned pointings cover most of the Northern lobe (aside from the Northern polar cap, where it lacked reference coverage, at the time) and Gattini planned pointings cover the entire Northern lobe, including the polar cap. The dialogue box shown in the righthand side portion of the figure displays the right ascension, declination, DECam field id, and DECam reference depth in different filters for a given DECam tile intersecting with the localization.}
    \label{fig:event_skymap}
\end{figure*}

Here, we highlight the importance of our ToO Marshal in handling multi-telescope scheduling from a common platform. This enables efficient and coordinated planning for conducting observations within the skymap, where the observation plans from different telescopes can complement one another. We demonstrate examples of complementary telescope follow-up between DECam, ZTF, Gattini-IR, and GROWTH-India in Figures \ref{fig:event_skymap} and \ref{fig:globe_plot}. 

\begin{figure*}[t]
    \centering
    \includegraphics[width=6cm, height=8cm]{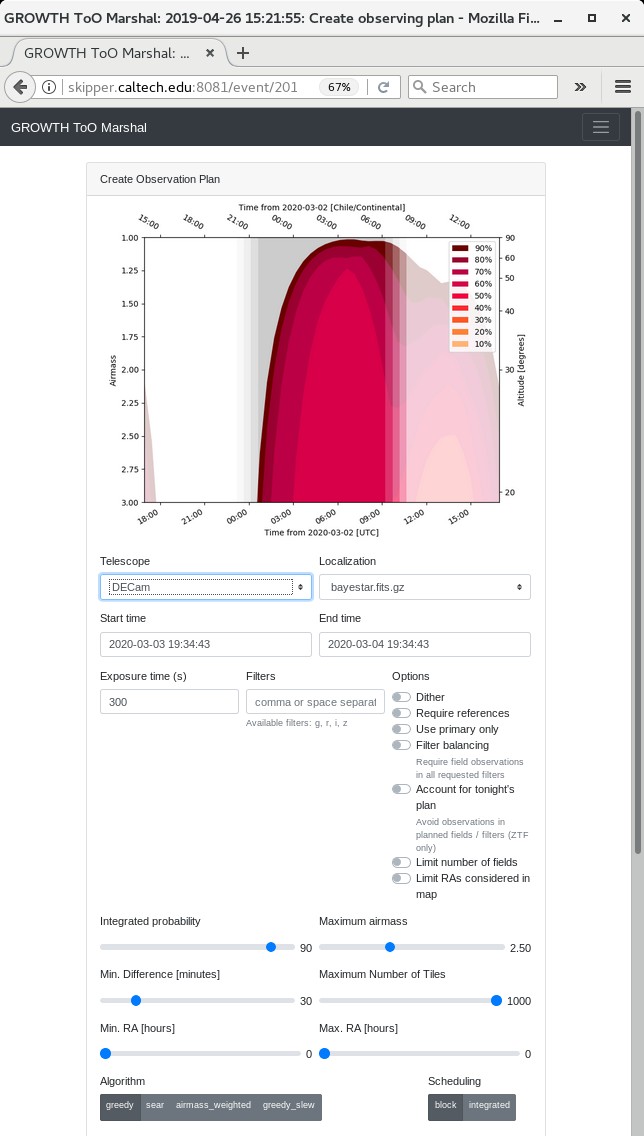}
    \caption{A screenshot of our planning page, used to create customized observation plans for each skymap and telescope. Some scheduling features we highlight here,that we have frequently used in our follow-up observations, include the greedy-slew algorithm that down-weights large slews when scheduling observations, filter balancing, to facilitate multi-epoch scheduling of the skymap, the ability to slice skymaps by right ascension, for better handling of disjointed skymaps. These features, as well as a few others, are described in more detail in AlMualla et al. 2020, in prep.}
    \label{fig:plan_new}
\end{figure*}

\begin{figure*}
    \centering
    \includegraphics[width=0.4\textwidth]{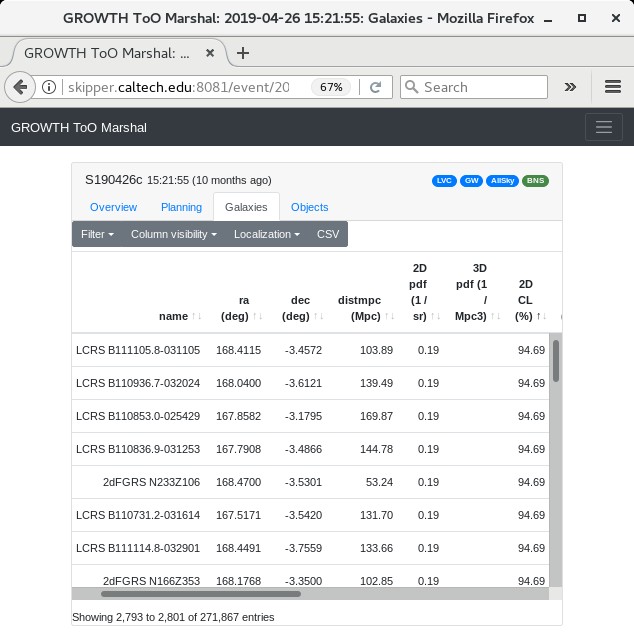}\includegraphics[width=0.4\textwidth]{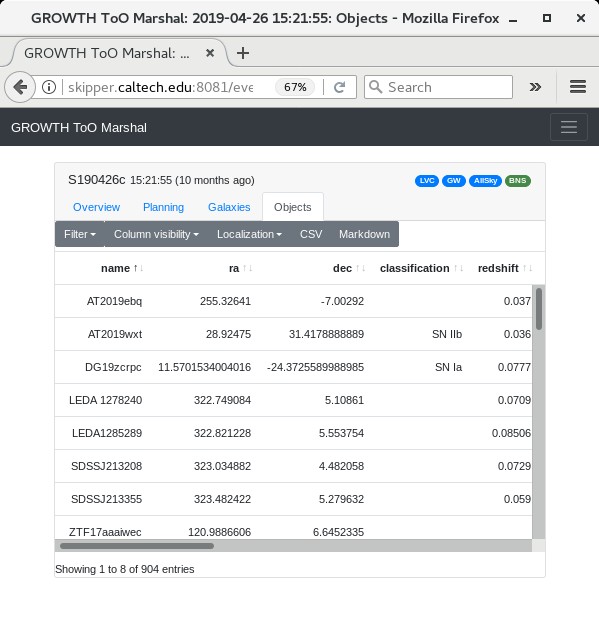}
    \caption{Screenshots from the GROWTH ToO Marshal, displaying our galaxies and objects pages for S190426c. Left: table containing eight galaxies falling within the 90\% credible volume of the 3D skymap; Right: table containing a subset of the objects falling within the 90\% credible region of the 2D skymap.}
    \label{fig:galaxies_objects}
\end{figure*}

\begin{figure}
    \centering
    \includegraphics[width=0.75\textwidth]{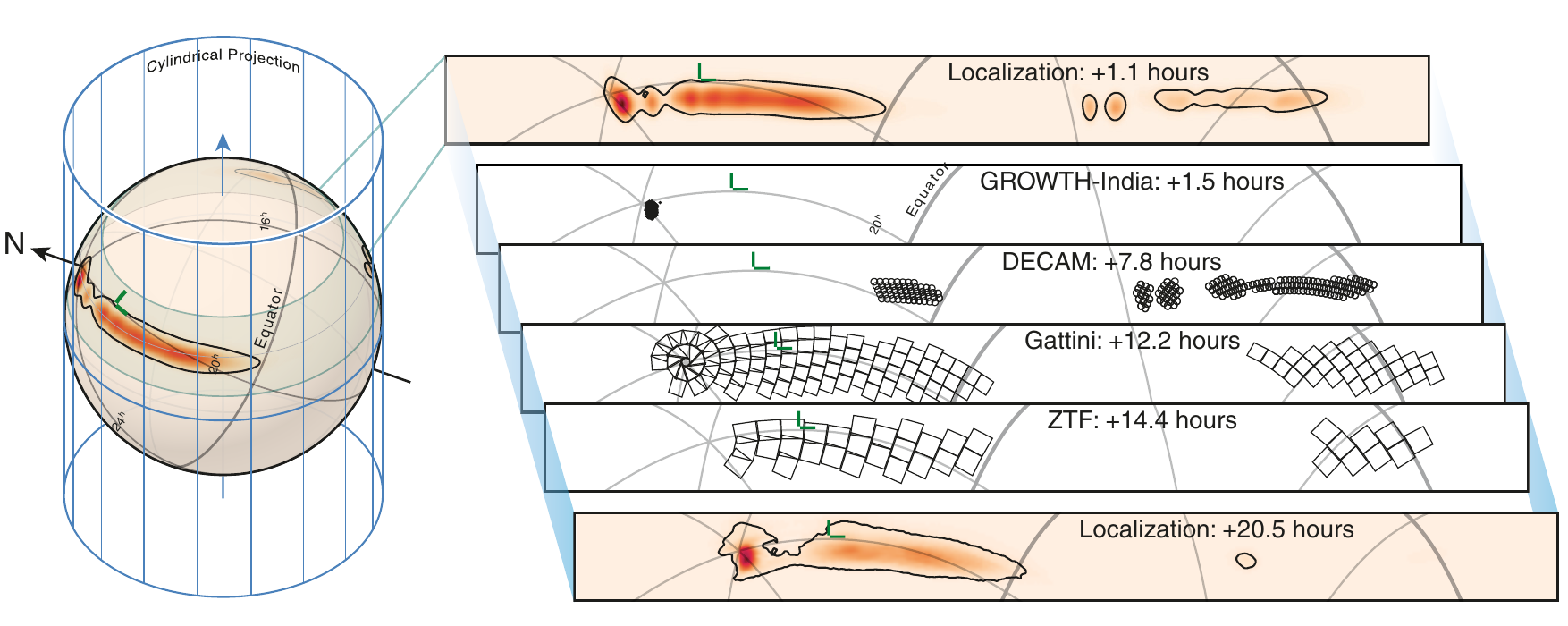}
    \caption{Follow-ups performed with ZTF, Palomar Gattini-IR, DECam, and GROWTH-India as a part of the GROWTH network during S190426c. In contrast to Figure \ref{fig:event_skymap} which shows an example of the planned pointings with the GROWTH telescope network, this figure highlights our actual completed observations within the skymap of S190426c. The lefthand side shows a globe with the skymap of S190426c projected onto it. The inset panels, from top to bottom, show 1) the initial BAYESTAR localization, 1.1 hours after the merger, followed by 2) a tiled coverage of the highest probability patch in the skymap by the GROWTH-India telescope in the $r$-band to a depth of $r<20.6$ 3) observations by Gattini-IR of most of the Northern portion of the skymap to J$<16.2$, 4) DECam coverage of the bottom portion of the Northern lobe, and Southern lobe to $r<23.4$, $z<22.5$, 5) ZTF observations of parts of the Northern and Southern lobe to a depth of $g<21.7$, $r<21.7$. Unfortunately, the localization shifted the probability away from the regions observed by DECam 20.5 hours after the merger, as shown as inset panel 6.}
    \label{fig:globe_plot}
\end{figure}


\subsection{DECam in the GROWTH network} \label{sec:decam_growth}
The Dark Energy Camera (DECam, \citealt{FlDi2015}) is a 3 sq. deg. FOV imager mounted on the 4m-class Blanco telescope, a member of the Cerro Tololo Inter-American Observatories in Chile. With its relatively wide field of view and high sensitivity, DECam is an excellent instrument for GW follow-up observations, as demonstrated during LIGO/Virgo O2 \citep{SoKe2016, CoBe2016} which culminated with GW170817 \citep{SoHo2017, CoBe2017}.

In 2018 and 2019 our DECam-GROWTH team were granted time for target-of-opportunity (ToO) observations for the follow-up of BNS and NSBH events detected by LIGO (PIs Igor Andreoni and Danny Goldstein; ``Public DECam Follow-up of Neutron Star Mergers During O3"; NOAO proposal IDs 2018B-0942, 2019A-0205, 2019B-0353). We specifically requested that our data be immediately made public after it was taken. As an outcome of the time allocation process, data acquired by all teams who were granted ToO triggers with DECam for the follow-up of GW events were made immediately public during O3.

 While ZTF is capable of mapping sky regions spanning thousands of square degrees in tens of pointings due to its massive footprint, its limiting magnitude hinders ZTF from achieving the necessary depth to follow-up distant GW events, and the limited filter coverage ($g$ and $r$, in O3a) are not the most optimal for the discovery of early-time kilonova emission. DECam, while suited for searches spanning a few hundreds of square degrees with its field-of-view and slew rate, is primed for following up the most distant GW events as it is capable of achieving a 3$\sigma$ magnitude of $r>22.3$ in 40 second exposures, as demonstrated in \citealt{AnGo2019}. The wider range of filters it offers (including $g$, $r$, $i$, and $z$) compared to ZTF can contribute towards capturing the intra- and inter-night color evolution of a kilonova \citep{FlDi2015}. By coordinating observations between DECam and ZTF using methods described in \citep{CoAn2019}, the GROWTH team can continue to map very large sky regions with ZTF while performing deep searches within the highest probability region with DECam. 


\section{DECam-GROWTH follow-up of O3a events} \label{sec:followup}
During the first half of O3, we triggered the DECam-GROWTH program twice for the follow-up of S190426c and S190510g. For other two events, S190728q and S190814bv, DECam was triggered and operated by the DESGW team; we ran our image-processing pipeline and conducted analyses based on the follow-ups. In these proceedings, we do not discuss S190728q as it was officially classified as a binary black hole, and is unlikely to have an associated EM counterpart. We describe the remaining three on an event-by-event basis, below. As described in Sec. \ref{sec:growth}, all of our triggered observations were planned and executed using the GROWTH ToO Marshal. Our nominal observing strategy for following up BNS localized to within 150 sq. deg. and within 200 Mpc is to conduct three epochs consisting of $g$-$z$-$g$ observations on the first night, and $g$-$z$ observations on the second night. This filter combination is primed to detect the fast blue optical emission expected for a GW170817-like kilonova and to single out a kilonova based on its rapid color evolution.  We performed image differencing for all of these follow-ups using an automated real-time image processing and image subtraction pipeline \citep{GoAn2019}. The pipeline launches a c5.18xlarge spot EC2 instance with 72 vCPUs and 144 GB of RAM for every exposure. The pipeline astrometrically and photometrically calibrates images in parallel, creates references, performs image subtractions, identifies candidates, filters them using \texttt{autoScan} \citep{GoDA2015} and performs aperture photometry. Each exposure takes $\sim20$ minutes to process. This image processing pipeline, the central engine powering our searches, runs now on Amazon Web Services Elastic Compute Cloud (EC2); the results of the pipeline are stored on Amazon Simple Storage Service (S3). A more complete and extensive description of the image processing and subtraction pipeline can be found in \citealt{GoAn2019}.
The reference templates for the imaged regions were drawn from three publicly available DECam datasets: Dark Energy Survey DR1 \citep{DES2016, Ab2018}, DECam Legacy Survey \citep{Dey2019}, and BLanco Imaging of the Southern Sky (BLISS; \citealt{SoAn2017}).  Table \ref{tab:summary} summarizes the follow-ups performed with DECam via the DECam-GROWTH and DESGW programs \citep{GoAn2019, AnGo2019, gcn25336, gcn25373, gcn25398, gcn25425, gcn25486, AnGo2020}.

\begin{table*}
    \label{tab:summary}
    \centering
    \begin{tabular}{cccccccc}
    \hline\hline
        event & classification & loc. area & distance & coverage & prob & depth & p$_{terrestrial}$\\
        \hline
        S190426c & NSBH & 1131 & 377 $\pm$ 100 & 525 & 0.08 & 22.9 & 0.14\\ 
        \hline
        S190510g & BNS & 1166 & 227 $\pm$ 92 & 249 & 0.62 & 23.1 & 0.58\\ 
        \hline
        S190728q & BBH & 104 & 874 $\pm$ 171 & 108 & 0.81 & 23.0 & 0.00\\
        \hline
        S190814bv & NSBH & 23 & 267 $\pm$ 52 & 64 & 0.98 & 21.0 & 0.00\\ 
    \hline\hline
    \end{tabular}
    \caption{Summary of the DECam-GROWTH (Goldstein and Andreoni et al., 2019b; Andreoni and Goldstein et al., 2019a,b) and DESGW (Soares-Santos et al. 2019a,b,c; Herner et al. 2019a,b)follow-up observations of LIGO detections during O3a.}
\end{table*}

\subsection{S190426c} 
(Details in \citealt{GoAn2019})
Initially classified as a binary neutron star merger, with descending probabilities of MassGap, neutron star-black hole (NSBH), and terrestrial, S190426c \citep{gcn24237} was of interest to our DECam-GROWTH program, as it was the first potential NSBH of O3. Part of the skymap, initially spanning 1262 sq. deg., was accessible with DECam. We generated a plan based on the greedy scheduling algorithm, with an integrated filter strategy that would conduct observations in $r$ and $z$, redder bands than the nominal strategy to detect the hypothesized emission from a NSBH merger.  Triggering observations starting at 2019-04-26 22:57:35 UT and ending at 2019-04-27 10:25:54 UT, we covered a total of 525 sq. deg. and 16\% probability in the BAYESTAR skymap to limiting magnitudes of 22.5 in $z$ and 22.9 in $r$. The updated LALInference skymap released the following night eliminated the probability in the regions we had already observed, so we did not trigger any further ToO observations. We identified a handful of candidates from our initial observations using the image subtraction pipeline that were systematically ruled out, or shown to be excluded from the updated LALInference map \citep{gcn24277}.

In Figure \ref{fig:globe_plot}, we highlight the importance of having a network of search telescopes in order to perform continuous and complementary follow-up of GW skymaps, aided in a large part by our GROWTH ToO Marshal. During our observational campaign of S190426c, we covered $>90\%$ of the probability in the skymap using the joint observations from four telescopes. The sequence of the observations was determined by the time at which the observable portions of the skymap was visible from each telescope site.

\subsection{S190510g} 
(Details in \citealt{AnGo2019})
Within a 15 day timespan, the LIGO+Virgo detector network observed another BNS merger candidate, S190510g, spanning 3462 sq. deg, with a distance of 269 $\pm$ 108 in the BAYESTAR skymap \citep{gcn24442}. Given the BNS classification, combined with the high distance, we triggered our DECam follow-up program. DECam imposes a strict engineering constraint on the hour angle of the telescope as a function of the declination; when attempting to address the hour angle limit, we discovered that scheduling observations using the default greedy algorithm introduced large slewing overheads into our plan. Thus we implemented a modified version of the greedy scheduling algorithm to down-weight large slews, and used this to schedule our observations of this event. We began our observations at 2019-05-10 06:00:25 UT, three hours after the merger, adopting a $g$-$z$ strategy for KNe from BNS mergers. Only one part of the skymap was above the horizon, as it was already Chilean night-time. Due to an unexpected system failure that night, we ran into the hour angle limit, and could only cover $\sim$15\% probability in $g$ and 10\% in $z$. Before our second night of observations, LIGO+Virgo released an updated LALInference skymap \citep{gcn24448}, shifting the probability from the two equatorial patches we observed to one patch at -30$^\circ$ declination, with the inner 50\% contour spanning only 31 sq. deg. Focusing on the highest probability patch, we conducted observations in $g$-$z$-$r$; since it was the second night after merger, the $g$-$z$ color combination would allow us to probe a fast blue KN component if it existed, while checking for redder emission detectable in the $r$-band. Employing this strategy, we covered a total of 67\% probability over three filter epochs. After our observations ended, the LVC re-classified this event as having 58\% terrestrial probability and the remainder BNS \citep{gcn24489}. Nevertheless, we employed our DECam image processing pipeline described earlier to identify candidates and reduced the set of over 150,000 transients to 12 viable counterpart candidates which we reported via GCN, although none of the candidates looked promising enough to be a KN. Comparing our upper limits against a set of existing kilonova models \citep{Bulla2019}, we demonstrate that with our observations, if this event was astrophysical, we would have detected a GW170817-like KN if it fell within the observed region at the median distance of this event, and at favorable viewing angles (See Figure 4 in \citealt{AnGo2019}).

\subsection{S190814bv} 
(Details in \citealt{AnGo2020})
With a 90\% localization of just 23 sq. deg. and an extremely low false alarm rate of $\sim$1/10$^{25}$ years, S190814bv is the most well-localized, high significance event amongst the BNS and NSBH merger candidates detected by LIGO to date. As its initial classification was MassGap, it fell under the purview of the DESGW program 2019B-0372 (PI: Soares-Santos; \citealt{SoKe2016}), rather than our DECam-GROWTH program. The DESGW team triggered ToO observations of DECam starting on 2019-08-15 06:32:43 UTC, and tiled the 98\% credible region 10 times over the course of 6 Chilean calendar nights (2019-08-14, 2019-08-15, 2019-08-16, 2019-08-17, 2019-08-20, 2019-08-30), increasing the searched depth with subsequent nights of observations. Observations were carried out in the $i$ and $z$ filters in order to capture the rapid reddening expected of a kilonova from a NSBH merger. During each night of observations planned and executed by the DESGW team, we retrieved the publicly available data and vetted candidates discovered by our automated image processing pipeline. Due to the 5$\sigma$ photometric limiting depth of the observations over the six nights ranging from $i$=21.1-24.1 and $z$=20.9-23.2, we limited our analysis to the $\pm2\sigma$ distances quoted for the event (Zhou et al., in prep). In practice, this involved employing a photometric redshift catalog created from the DESI Legacy Imaging Survey Data Release 8 \citep{Dey2019} to estimate the photometric redshifts of candidates with a host galaxy and reject candidates falling outside the $\pm 2\sigma$ LVC distance. We reported all candidates of interest identified during the vetting procedures via GCN. Our analysis yielded seven significant candidates of interest with initial discovery by DECam or ZTF, for which we obtained photometric and spectroscopic follow-up \citep{AnGo2020}; in addition to these, we followed up three candidates identified by DESGW and Pan-STARRS teams \citep{gcn25398, gcn25425, gcn25486, gcn25417}. The most promising KN counterpart candidate amongst these was DG19wxnjc/AT2019npv \citep{gcn25393}, whose initial detections in optical and NIR bands, reddening rate $\delta$($i$-$z) \sim$0.05 mags/day and subsequent non-detection 5$\sigma$ limit of $J$=21.4 made for a convincing case. On 2019-08-24 we obtained one NIR spectrum with Keck II+NIRES that appeared mostly featureless, with He I features at the redshift of the host galaxy, demonstrating that the candidate was a SN Ib/c (See Figure 2 of \citep{AnGo2020}); \citealt{gcn25483} confirmed the classification as a SN Ic.  The remaining significant candidates were either classified as supernovae or shown to have pre-detections in other surveys, excluding the possibility that they could be associated with S190814bv.

As a result of the well-cadenced, deep synoptic limits placed on the KN emission of a potential counterpart of S190814bv by these DECam observations, we were able to place significant constraints on the parameters of the merging system, if indeed there was a detectable optical/NIR countepart counterpart within the 23 sq. deg. localization. First, we consider 2D kilonova models obtained with the radiative transfer code \texttt{possis} \citep{Bulla2019} which are parameterized in terms of the ejecta mass M$_{ej}$, half-opening angle of the equatorial lanthanide-rich component $\phi$, and the viewing angle $\theta_{obs}$. We select $\phi=15^\circ$ and $\phi=30^\circ$ as the values for our ejecta half-opening angles, based on previous numerical simulations \citep{KaKy2016, FeTc2019}. For reference, the model best-matching with GW170817 has $\phi=30^\circ$, M$_{ej}=0.05\rm M_{sun}$, and $\cos{\theta_{obs}}=0.9$ \citep{DhBu2019}. In Figure 5 of \citealt{AnGo2020}, the top panel displays the lightcurve models that can be ruled out for these two opening angles based on the upper limits in $i$ and $z$ bands over the six nights of observations. The bottom panel of the figure demonstrates the regions of M$_{ej}$-$\theta_{obs}$ parameter space that are allowed and disallowed based on the non-detection upper limits.  The brighest kilonovae, with polar viewing angles (face-on, $\theta_{obs}$=0.0) are ruled out by the DECam observations for any distance assumption. However, only when assuming the closest distance, a polar viewing angle, and a half-opening angle of 15$^\circ$ can one place the most stringent constraints on ejecta mass (M$_{ej} <$0.05M$_{sun}$); for more conservative distances, viewing angles, and half-opening angles, we set the constraint that M$_{ej} < 0.10 \rm M_{sun}$. Overall, the limits are most constraining when a nearby distance (light blue) is assumed for the event; we see that deeper observations earlier on would have placed more stringent constraints on the kilonova parameters.

In Figure 6 of \citep{AnGo2019}, we assume a different kilonova model with spherical ejecta with a density profile $\rho \alpha v^{-n}$; the resulting kilonova models rely on a heating rate formalism described in \citealt{HoNa2019}. By using these models to constrain the $\kappa$-M$_{ej}$ (opacity - ejecta mass) space, we find that a KN associated with S190814bv cannot have a lanthanide-poor ejecta mass component that is $>0.05 (0.03) \rm M_{sun}$ if the merger occurred at 267 (215) Mpc. The shaded regions in the figure correspond to regions of parameter space where the KN lightcurve models are brighter than our DECam upper limits at the assumed distance. These results are roughly consistent with the results obtained using the \texttt{POSSIS} lightcurves, under favorable viewing angles.

Finally, synthesizing our constraints on the ejecta mass from both sets of KN models, we translate them to constraints on the merging binary system, contingent upon a set of assumptions. Figure 7 of \citealt{AnGo2020} displays the constraints on the maximum aligned spin for various combinations of mass ratio between the neutron star and black hole, and the tidal deformability of the neutron star, for M$_{ej}<0.03\rm M_{sun}$. Assuming 1) a dimensionless tidal deformability $\Lambda_{NS}>$800, which was the maximum tidal deformability for GW170817, where compactness of a star decreases with increasing $\Lambda_{NS}$, 2) M$_{ej}<0.03 \rm M_{sun}$, based on the KN models from \citealt{HoNa2019}, assuming the closest distance of 215 Mpc and 3) aligned spin between the neutron star and black hole, the data constrain the maximum aligned spin of the black hole to have $\chi_{BH}<0.7$ for mass ratios Q$>6$. 

We note that, considering the non-detection of a counterpart to S190814bv in the deep synoptic DECam follow-ups performed, we cannot exclude the possibility that S190814bv may also be a binary black hole merger or a NSBH merger system with a very large mass ratio, such that the NS plunged directly into the black hole.

\section{Summary}
Automatic telescope scheduling is crucial for the discovery of EM counterparts to GW sources. Here we presented the GROWTH ToO Marshal platform used to dynamically schedule multi-messenger follow-up with 5 telescopes, including ZTF, DECam, Gattini-IR, GROWTH-India, and KPED. DECam is a strong asset in contributing to the EM follow-ups of GW events in conjunction with the GROWTH network when searching skymaps spanning hundreds of square degrees, due to its location in the Southern hemisphere, its deep sensitivity, and multi-band coverage. However, DECam excels particularly in the follow-ups of distant, well-localized events, such as S190814bv; deep and targeted DECam follow-ups of S190814bv enabled us to place meaningful constraints on kilonova parameter space and binary properties, even given non-detection of a kilonova. 

This work was supported by the GROWTH project funded by the National Science Foundation under Grant No 1545949.

\bibliography{references}

\end{document}